\documentclass[12pt]{article}
\usepackage{epsf}
\usepackage{amsmath}
\usepackage{graphics}

\renewcommand{\thefootnote}{\#\arabic{footnote}}

\begin{document}

\setcounter{footnote}{0}
\begin{titlepage}

\begin{center}

\hfill hep-ph/0501007\\
\hfill TU-736\\
\hfill January 2005\\

\vskip .5in

{\Large \bf
Relaxing Constraints on Inflation Models\\
with Curvaton
}

\vskip .45in

{\large
Takeo Moroi$^a$, Tomo Takahashi$^b$ and Yoshikazu Toyoda$^a$
}

\vskip .45in

{\em
$^a$Department of Physics, Tohoku University\\
Sendai 980-8587, Japan \\
}

\vskip .2in

{\em
$^b$Institute for Cosmic Ray Research,
University of Tokyo\\
Kashiwa 277-8582, Japan
}

\end{center}

\vskip .4in

\begin{abstract}

We consider the effects of the curvaton, late-decaying scalar
condensation, to observational constraints on inflation models.  From
current observations of cosmic density fluctuations, severe
constraints on some class of inflation models are obtained, in
particular, on the chaotic inflation with higher-power monomials, the
natural inflation, and the new inflation.  We study how the curvaton
scenario changes (and relaxes) the constraints on these models.

\vspace{1cm}

\end{abstract}

\end{titlepage}

\renewcommand{\thepage}{\arabic{page}}
\setcounter{page}{1}
\renewcommand{\thefootnote}{\#\arabic{footnote}}
\renewcommand{\theequation}{\thesection.\arabic{equation}}

\section{Introduction}
\setcounter{equation}{0}

Inflation, a superluminal expansion at the early stage of the universe
\cite{inflation}, is one of the most promising ideas to solve the
flatness and horizon problems which are serious drawbacks of the
standard hot big bang model.  Inflation can be caused by the potential
energy of a scalar field called inflaton which slowly rolls down the
potential.  Since quantum fluctuation of the inflaton during inflation
is usually assumed to be the origin of cosmic density perturbation,
models of inflation can be tested by observations of cosmic microwave
background (CMB), large scale structure and so on.  Recent results
from Wilkinson microwave anisotropy probe (WMAP) \cite{Bennett:2003bz}
gave severe constraints on inflation models \cite{Peiris:2003ff}.
For example, chaotic inflation model \cite{Linde:1983gd} with quartic
potential $V_{\rm inf}\propto\chi^4$, where $\chi$ is the inflaton, is
on verge of exclusion by observations assuming the standard thermal
history of the universe.

However today's cosmic fluctuation can originate to some source other
than the inflaton.  In the curvaton scenario \cite{curvaton},
adiabatic fluctuation can be given by primordial fluctuation of some
late-decaying scalar field which is called curvaton.  In this
scenario, the superluminal expansion is caused by the inflaton while
(some part of) the fluctuation responsible for the cosmic fluctuation
today is provided by the curvaton field.  Thus, in this case,
constraints on models of inflation can be alleviated to some extent.
Even if the fluctuation of the inflaton is partly relevant to the
today's fluctuation, we can expect that constraints on models of
inflation can be relaxed compared to the case without the curvaton.
The authors of Ref.~\cite{Langlois:2004nn} studied the constraint on
the quartic chaotic inflation model in the curvaton scenario and
showed that the quartic chaotic inflation model is still viable in the
curvaton scenario.

In this paper, we investigate this issue in detail for various
inflation models which, in particular are on verge or already excluded
in the standard inflationary scenario.  We will show that some
inflation models can be made viable in the curvaton scenario.

The organization of this paper is as follows. In the next section, we
give a brief review of the scenario where both inflaton and curvaton
fluctuations give today's cosmic fluctuations and define our
notations. In addition, we also describe our analysis method. In
Section \ref{sec:results}, we discuss to what extent constraints on
inflation models are alleviated in the curvaton scenario starting from
the chaotic inflation model, the natural inflation model and the new
inflation model.  The conclusion of our study is given in the final
section.

\section{Formalism}
\setcounter{equation}{0}

\subsection{Background Evolution}
The aim of this paper is to investigate to what extent constraints on
inflation models can be relaxed in the curvaton scenario.  Thus we
need to discuss density perturbations.  However, before discussing
density perturbations, first we discuss the background evolution and
fix our notations.

We start with the inflationary era. During inflation, there assumed to
be two scalar fields, one is the inflaton $\chi$ and the other is the
curvaton $\phi$ which follow the following equations of motion;
\begin{eqnarray}
  \ddot{\chi} + 3 H \dot{\chi} + \frac{dV_{\rm inf}}{d\chi} = 0, 
  \label{eq:EOMchi} \\
  \ddot{\phi} + 3 H \dot{\phi} + \frac{dV_{\rm curv}}{d\phi} = 0,
  \label{eq:EOMphi}
\end{eqnarray}
where the dot represents derivative with respect to time, $V_{\rm inf}$
and $V_{\rm curv}$ are the potentials for the inflaton and the
curvaton, respectively. $H$ is the Hubble parameter which is given by
\begin{equation}
  H^2 = \left( \frac{\dot{a}}{a} \right)^2 
  = \frac{1}{3 M_{\rm pl}^2} \rho_{\rm tot},
\end{equation}
where $a$ is the scale factor, $M_{\rm pl}\simeq 2.4\times 10^{18}\ 
{\rm GeV}$ is the reduced Planck scale, and $\rho_{\rm tot}$ the total
energy density of the universe.  During inflation, it is given by the
sum of contributions from the inflaton and the curvaton as
\begin{equation}
\rho_{\rm tot} = \rho_\chi + \rho_\phi 
= \frac{1}{2} \dot{\chi}^2  + V_{\rm inf}
+ \frac{1}{2} \dot{\phi}^2 + V_{\rm curv}.
\end{equation}
For the potential of the inflaton, we consider various possibilities
which we will describe later.  For the curvaton, for simplicity, we
consider the quadratic potential in this paper;
\begin{equation}
  V(\phi) = \frac{1}{2} m_\phi^2 \phi^2,
  \label{eq:Vphi}
\end{equation}
where $m_\phi$ is the mass of the curvaton.

After the inflation, the inflaton begins to oscillate around the
minimum of the potential.  We call this epoch as $\chi$-dominated or
$\chi$D era.  Then, the inflaton decays when $H\sim\Gamma_{\chi}$
(with $\Gamma_\chi$ being the decay rate of the inflaton).  In this
period, the evolution of the inflaton is described by Eq.\
(\ref{eq:EOMchi}) with the decay term being included.  Once the
expansion rate $H$ becomes much smaller than the oscillation frequency
of $\chi$, however, it is rather convenient to consider the averaged
value of the energy density of $\chi$ and to follow its evolution.
Indeed, for the inflaton potential $V_{\rm inf}\propto\chi^n$,
$\rho_\chi$ follows the equation
\begin{equation}
  \dot{\rho}_\chi = - 3(1+w_\chi) H \rho_\chi - \Gamma_\chi \rho_\chi,
  \label{eq:dot(rho_chi)}
\end{equation}
with $w_\chi=(n-2)/(n+2)$.  Thus, the energy density of the
oscillating inflaton behaves as \cite{Turner:1983he}
\begin{equation}
  \rho_\chi \propto a^{-\frac{6n}{n+2}}.
\end{equation}
In our study, we only consider the case where $\Gamma_\chi$ is small
enough so that the inflaton decays when $H$ is much smaller than the
oscillation frequency of $\chi$.  Thus, the reheating processes by the
inflaton is studied by using Eq.\ (\ref{eq:dot(rho_chi)}).  
In addition, the energy density of the radiation evolves as
\begin{eqnarray}
  \dot{\rho}_{\rm rad}  =
  - 4 H \rho_{\rm rad} +  \Gamma_\chi \rho_\chi
  +  \Gamma_\phi \rho_\phi,
\end{eqnarray}
with $\Gamma_\phi$ being the decay rate of the curvaton.  Here, we
included the effect of the curvaton decay, which will become inportant
in the following discussion.  When $H\sim\Gamma_{\chi}$, the inflaton
decays into radiation, then the universe is reheated to become
radiation-dominated. We call this radiation dominated era ``RD1.''

When the expansion rate of the universe becomes comparable to
$m_\phi$, the curvaton also starts to oscillate, then it behaves like
matter component for the potential given in Eq.~(\ref{eq:Vphi}).
Thus, at some epoch the universe becomes curvaton dominated. We call
this epoch ``$\phi$D'' era. After the $\phi$D era, the curvaton decays
into radiation then the universe becomes radiation dominated again.
We call this epoch ``RD2'' era. The time when the RD2 era starts
depends on the decay rate of the curvaton $\Gamma_\phi$.  From $\phi$D
era to RD2 era, the universe consists of radiation and the oscillating
curvaton field. Since the curvaton behaves like matter in this epoch,
energy density of the curvaton is governed by the following equation
\begin{eqnarray}
  \dot{\rho_\phi}  &=& - 3 H \rho_\phi  -\Gamma_\phi \rho_\phi.
\end{eqnarray}

When the initial amplitude of the curvaton is as large as $\phi_{\rm
init} \sim M_{\rm pl}$, the curvaton can cause the second inflation
after the usual inflation provided by the inflaton.  In this case, the
second inflation epoch exists before $\phi$D era mentioned above.
Depending on the initial amplitude and mass of the curvaton, the
second inflation era can start during $\chi$D era or RD1 era.  Such
effect is also taken into account in our analysis (in particular, in
our numerical study to evaluate the $e$-folding number during the
inflation).

Here we summarize what we need to fix the background evolution.  As
for the background, the following parameters are necessary to
determine the thermal history of the universe except the potential of
the inflaton (and the initial amplitude of the inflaton): the mass of
the curvaton $m_\phi$, the initial amplitude of the curvaton
$\phi_{\rm init}$, the decay rates of the inflaton and the curvaton
$\Gamma_\chi$ and $\Gamma_\phi$.

\subsection{Density Perturbation}
Next we discuss the density perturbation in the scenario.  Since there
are two sources of cosmic density fluctuation in this case (i.e.,
the inflaton and the curvaton), we need to take account of the effects
of both of them.  Since the primordial fluctuations of the inflaton
and the curvaton are uncorrelated, we can study their effects
separately.  First we summarize the results for the standard case,
i.e., the case with the inflaton only. Then we will discuss how the
situation is modified in the curvaton scenario.

The quantum fluctuation of the inflaton $\delta\chi$ during inflation
generates the curvature perturbation as $\mathcal{R} = -
(H/\dot{\chi}) \delta \chi$. Since $\delta\chi\simeq H/2 \pi$, the
power spectrum of the curvature perturbation from the inflaton
fluctuation is
\begin{equation}
  P_\mathcal{R}^{\rm (inf)} = 
  \frac{1}{12 \pi^2 M_{\rm pl}^6} 
  \frac{V_{\rm inf}^3}{V_{\rm inf}^{'2}} \biggr|_{k=aH},
\end{equation}
where the ``prime'' is the derivative with respect to $\chi$.  Here,
we used the slow-roll approximation.  To discuss observational
consequences, we have to set up the initial condition during radiation
dominated era after the decay of the inflaton.  For this purpose, we
represent the primordial power spectrum with the (Bardeen's)
gravitational potential $\Phi$ which appears in the perturbed metric
in the conformal Newtonian (or longitudinal) gauge as
\begin{equation}
  ds^2 = -a^2( 1 + 2\Phi) d\tau^2 + a^2 ( 1 - 2\Psi) dx^2,
\end{equation}
with $\tau$ being the conformal time.  During radiation-dominated era,
$\Phi$ and $\mathcal{R}$ are related as $\Phi = -(2/3)
\mathcal{R}$. Thus the power spectrum of $\Phi$ is given by $P_\Phi =
(4/9) P_\mathcal{R}$.  The spectral index of the primordial power
spectrum is
\begin{equation}
n_s -1 \equiv \frac{d \ln P_\Phi}{d \ln k}. 
\end{equation}
Using the slow-roll parameters defined as\footnote
{In this paper, we use so-called ``potential'' slow-roll parameters
which are defined using the value of the potential not using the
Hubble parameter.}
\begin{eqnarray}
  \epsilon \equiv \frac{1}{2} M_{\rm pl}^2 
  \left( \frac{V_{\rm inf}'}{V_{\rm inf}} \right)^2,
  ~~~
  \eta \equiv M_{\rm pl}^2 \frac{V_{\rm inf}''}{V_{\rm inf}},
\end{eqnarray}
the spectral index can be written as
\begin{equation}
n_s^{\rm (inf)} - 1 =  - 6 \epsilon + 2 \eta.
\label{eq:n_s}
\end{equation}

During inflation, the gravity wave can also be generated.  The
primordial gravity wave (tensor) power spectrum is given by
\begin{equation}
  P_T^{\rm (inf)}  = \frac{2V_{\rm inf}}{3 \pi^2 M_{\rm pl}^4}.
\end{equation}
With this expression, the tensor-scalar ratio $r$ is defined and given
by
\begin{equation}
  r^{\rm (inf)}
  \equiv 
  \frac{P_T^{\rm (inf)}}{P_\mathcal{R}^{\rm (inf)}}
  = 
  \frac{4}{9} \frac{P_T^{\rm (inf)}}{P_\Phi^{\rm (inf)}} 
  = 16 \epsilon.
\end{equation}

Now, we consider the fluctuations generated by the curvaton
fluctuation.  Let us start with summarizing the equations relevant for
the following discussion.  From the Einstein equation, we obtain the
perturbation equation for the metric perturbations:
\begin{equation}
  -k^2 \Phi 
  = 
  \frac{3}{2} \mathcal{H}^2 
  \left[ 
    \delta_{\rm tot} 
    + \frac{3\mathcal{H}}{k} (1 +\omega_{\rm tot} ) V_{\rm tot} 
    \right],
  \label{eq:poisson}
\end{equation}
where $\mathcal{H} = (1/a)(da/ d \tau)$ is the conformal Hubble
parameter, $\delta_X = \delta \rho_X/\rho_X$ and $V_X$ are density
perturbation and velocity perturbation of a component $X$. In the
above equation, ``tot'' denotes the total matter, and $w_{\rm
tot}=p_{\rm tot}/\rho_{\rm tot}$ is the equation-of-state parameter of
the total matter.  Assuming that the anisotropic stress is negligible,
the metric perturbation variables are related as $\Phi = \Psi$.  The
equations for the density and velocity perturbations of the component
whose equation-of-state parameter is $\omega_X$ are
\begin{eqnarray}
  \frac{d \delta_X}{d \tau} 
  &=& 
  - (1+\omega_X) \left( k V_X - 3 \frac{d \Psi}{d\tau} \right),
  \label{eq:delta_X}
  \\
  \frac{d V_X}{d \tau} 
  &=& 
  - \mathcal{H} ( 1 - 3 \omega_X )V_X 
  + \frac{\omega_X}{1+\omega_X} k\delta_X + k \Phi.
  \label{eq:V_X}
\end{eqnarray}

Evolutions of the fluctuations are followed by solving the above
equations (as well as those for the background quantities).  Since
there are two independent contributions to the cosmic density
fluctuations, i.e., the primordial fluctuations of the inflaton and
the curvaton, $\Phi_{\rm RD2}$ in this case has two terms which are
proportional to $\delta\chi_{\rm init}$ and $\delta\phi_{\rm init}$,
respectively:
\begin{equation}
  \Phi_{\rm RD2}
  = 
  - \frac{2}{3M_{\rm pl}^2} \frac{V_{\rm inf}}{V_{\rm inf}'}  
  \delta \chi_{\rm init}
  - f(X) \frac{\delta \phi_{\rm init}}{M_{\rm pl} },
  \label{eq:Phi_mod}
\end{equation}
where $\delta\phi_{\rm init}$ is the primordial fluctuation of the
curvaton, and
\begin{eqnarray}
  X = \frac{\phi_{\rm init}}{M_{\rm pl}}.
\end{eqnarray}
The first term of the right-hand side of Eq.\ (\ref{eq:Phi_mod}) is
the inflaton contribution which we have already discussed before.  The
second term is the curvaton contribution; in our study, $f(X)$ is
calculated by numerically solving the Einstein and Boltzmann equations.
Here, the curvaton contribution is parameterized by the function
$f(X)$ \cite{Langlois:2004nn}.  Although we have used the numerical
method to calculate $f(X)$, effects of the curvaton can be easily
evaluated for the cases where the initial amplitude of the curvaton is
much larger or much smaller than $M_{\rm pl}$
\cite{Langlois:2004nn,Moroi:2002rd}.  Indeed, for these cases, $f(X)$
is given by
\begin{eqnarray}
  f(X) \simeq \left\{
    \begin{array}{ll}
      \displaystyle{ \frac{4}{9X}} 
      & ~~~:~~~ \phi_{\rm init} \ll M_{\rm pl}
      \\ \\
      \displaystyle{ \frac{1}{3}X}
      & ~~~:~~~ \phi_{\rm init} \gg M_{\rm pl}
    \end{array} \right. .
\end{eqnarray}

Assuming that the inflaton and the curvaton are uncorrelated, we find
the power spectrum of $\Phi$ as \cite{Langlois:2004nn}
\begin{equation}
  P_\Phi = 
  \left[
  1 + \tilde{f}^2 (X) \epsilon 
  \right]
  \frac{V_{\rm inf}}{54 \pi^2 M_{\rm pl}^4 \epsilon},
\end{equation}
where $\tilde{f} = (3/\sqrt{2}) f$.
The scalar spectral index is given by using Eq.~(\ref{eq:n_s}) 
\begin{equation}
  n_s-1 =  
  - 2 \epsilon 
  +  \frac{ 2 \eta - 4 \epsilon}{1 + \tilde{f}^2 \epsilon} .
  \label{eq:ns_mod}
\end{equation}
The tensor power spectrum is not modified even with the
curvaton. However, since the scalar perturbation spectrum is modified,
the tensor-scalar ratio becomes
\begin{equation}
  r = \frac{16 \epsilon} {1 + \tilde{f}^2 \epsilon}.
  \label{eq:r_mod}
\end{equation}

\section{Results}
\label{sec:results}
\setcounter{equation}{0}

Now, we are ready to discuss how the constraints on inflation models
are alleviated with curvaton.  We investigate the following inflation
models; chaotic inflation models for several monomials, the natural
inflation model and the new inflation model.  For these models,
observations of CMB anisotropy and large scale structure provide
severe constraints and some part of the parameter space in these
models is excluded in the standard scenarios.

Including the curvaton contributions, we analyze the constraints on
these inflation models, making use of the observational constraints on
the scalar spectral index $n_s$ and the tensor-scalar ratio $r$.  In
our analysis, we first calculate $n_s$ and $r$ for a given model.  For
this purpose, it is important to determine the amplitude of the
inflaton field at the time of the horizon exit $\chi_*$; once $\chi_*$
is given, $n_s$ and $r$ are calculated by using Eqs.\
(\ref{eq:ns_mod}) and (\ref{eq:r_mod}).  (In this paper, the subscript
``$*$'' is used for the quantities at the time of the horizon exit.)
Importantly, $\chi_*$ depends on various parameters.  In order for the
systematic and precise determination of $\chi_*$, we followed the
evolution of the inflaton and the curvaton field (as well as the
energy density of radiation) from the inflationary era to the RD2 era
by numerically solving the Einstein and Boltzmann equations given in
the previous section.  Since we evaluate the values of $n_s$ and $r$
at $k = 0.01 {\rm Mpc}^{-1}$ to compare a model with observations of
CMB and large scale structure, $\chi_*$ is determined by the condition
$ H_* (a_*/a_0) = 0.01 {\rm Mpc}^{-1}, $ where $a_*$ and
$H_*$ are the scale factor and the expansion rate at $\chi=\chi_*$,
respectively, and $a_0$ is the present scale factor.  In our analysis,
$\chi_*$ is calculated as a function of $X$, $m_\phi$, and other model
parameters (in particular, the decay rates of the inflaton and
curvaton).  With $\chi_*$, we calculate $n_s$ and $r$.\footnote
{Strictly speaking, as for CMB data, we have to compare the data with
$C_l = C_l^{(\delta \chi)} + C_l^{(\delta \phi)}$ where $C_l^{(\delta
\chi)}$ and $C_l^{(\delta \phi)}$ are contributions from the
primordial fluctuation of the inflaton and the curvaton, respectively,
and similarly for observations of large scale structure. We checked
that calculations of $C_l$ using of the effective spectral index does
not affect our results (the final $C_l$ does not change within 0.1 \%
error).  Thus, for the purpose of this paper, we can use the effective
spectral index in our analysis.}
Then, we compare them with the observational constraints on $n_s$ and
$r$ obtained by many authors (see e.g., \cite{Leach:2003us}).  In our
analysis, allowed region on the $m_\phi$ vs.\ $X$ plane is given by
the region where $(n_s, r)$ calculated with $(X, m_\phi)$ falls onto
the 95 \% C.L. allowed region on the $n_s$ vs.\ $r$ plane given in
Ref.\ \cite{Leach:2003us}. In the cases of the natural inflation and
the new inflation, we show the constraints in other planes instead of
the $m_\phi$ vs. $X$ plane replacing $m_\phi$ with another parameter
in the inflaton sector. However, even in such a case, the analysis
method is the same as above.

\subsection{Chaotic Inflation}

We parameterize the potential for the chaotic inflation as
\begin{equation}
  V_{\rm inf} = 
  \lambda M_{\rm pl}^4 \left( \frac{\chi}{M_{\rm pl}} \right)^\alpha.
\end{equation}
Here, we require $\alpha$ to be an even integer for the positivity and
smoothness of the inflaton potential.  We consider the cases with
$\alpha\geq 4$, for which observations of CMB and large scale
structure provide severe constraints for the case without the
curvaton.

The slow-roll parameters in this model are given by
\begin{equation}
  \epsilon = \frac{1}{2} \alpha^2 \frac{M_{\rm pl}^2}{\chi_*^2},
  ~~~ 
  \eta = \alpha ( \alpha -1) \frac{M_{\rm pl}^2}{\chi_*^2}.
\end{equation}
The moment when the present horizon scale exits the horizon during
inflation is usually denoted with the $e$-folding number of inflation
which is defined as the logarithm of the ratio of the scale factors at
the horizon crossing and the end of inflation: $N_e \equiv \ln( a_*/
a_{\rm end})$.  We can write this quantity with the amplitude of the
inflaton
\begin{equation}
  N_e =
  \frac{1}{M_{\rm pl}^2} \int^{\chi_*}_{\chi_{\rm end}}  
  \frac{V_{\rm inf}}{V_{\rm inf}'} d \chi,
\end{equation}
where $\chi_{\rm end}$ is determined as the inflaton amplitude when
one of the slow-roll parameters becomes 1.

Since the amplitude of the inflaton at the end of inflation is much
smaller than $\chi_*$, the $e$-folding number can be written as
\begin{equation}
  N_e \simeq \frac{1}{2 \alpha M_{\rm pl}^2} \chi_*^2.
\end{equation}
Although $N_e$ depends on the thermal history of the universe, it is
often assumed that $N_e\sim 50$. If we take $N_e=50$, the scalar
spectral index and the tensor-scalar ratio $(n_s-1,r)$ are $(-0.06,
0.32)$ for $\alpha = 4$, which are marginally excluded by the WMAP
result. For $\alpha =6$ and 8, $(n_s -1, r) = (-0.08, 0.48)$ and
$(-0.1, 0.64)$, respectively, which are completely ruled out.

However, in the curvaton scenario, fluctuation of the curvaton also
affects the primordial spectrum.  As for the spectral index, we can
see from Eq.~(\ref{eq:ns_mod}) that $n_s$ becomes smaller in the case
with curvaton.  The tensor-scalar ratio also becomes smaller as we can
see from Eq.~(\ref{eq:r_mod}).  Thus constraints on the inflation
models based on $n_s$ and $r$ can be relaxed with curvaton.  However,
notice that, in the curvaton scenario, the background evolution is
also changed.  In particular, when the second inflation occurs, it may
decrease the $e$-foldings during the first inflation by $20-30$.  Thus
$N_e$ during inflation changes depending on the parameters of the
curvaton such as mass, initial amplitude and decay rate.  (Of course
it depends on the inflaton parameters such as the decay rate of the
inflaton.)  This also changes the constraints on inflation models.
Details of the change will be discussed for each models.

\subsubsection{$\alpha=4$ case}

The chaotic inflation with $\alpha=4$ is on verge of exclusion by
cosmological observations as mentioned above.  Here we consider how
this constraint changes in the curvaton scenario.

First we describe how we treat free parameters in this model.  The
free parameters in this model are the coupling in the potential
$\lambda$, the mass of the curvaton $m_\phi$, the initial amplitude of
the curvaton $\phi_{\rm init}$ and the decay rate of the curvaton
$\Gamma_\phi$. Although the decay rate of the inflaton also affect the
number of $e$-foldings in general, oscillating inflaton field in this
model behaves as radiation, namely the energy density of the
oscillating scalar field is given by $\rho_\chi \propto a^{-4}$. Hence
$\Gamma_\chi$ is irrelevant to determine the number of $e$-foldings in
this model. For the coupling parameter $\lambda$, it can be fixed to
have right amount of density fluctuation.  However, the scalar
spectral index and the tensor-scalar ratio are independent of
$\lambda$, as we can see from Eq.~(\ref{eq:ns_mod}).  Since we
discriminate the model using the constraints on $n_s$ and $r$, we do
not discuss $\lambda$ in the following.

\begin{figure}[t]
\begin{center}
\scalebox{1}{\includegraphics{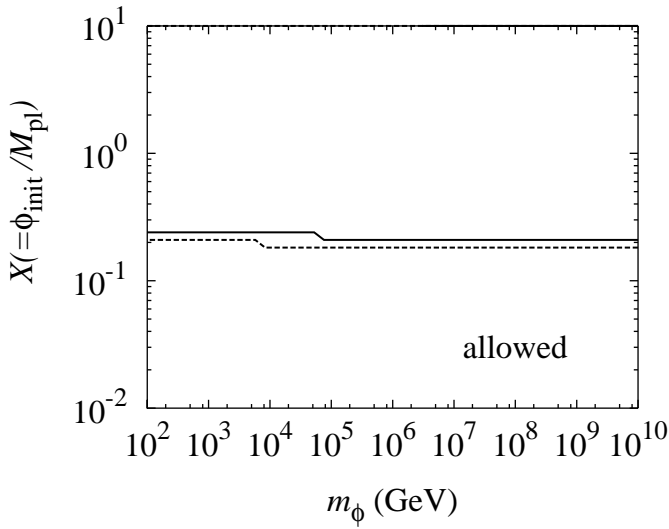}}
\caption{The 95 \% C.L. allowed regions are shown for $\alpha=4$.  The
decay rate of the curvaton is assumed to be $\Gamma_\phi = 10^{-18}$
GeV (solid line) and $\Gamma_\phi = 10^{-10}$ GeV (dashed line). Lower
side of the horizontal line is allowed for the given $\Gamma_\phi$. }
\label{fig:phi4}
\end{center}
\end{figure}

\begin{figure}[t]
\begin{center}
\includegraphics{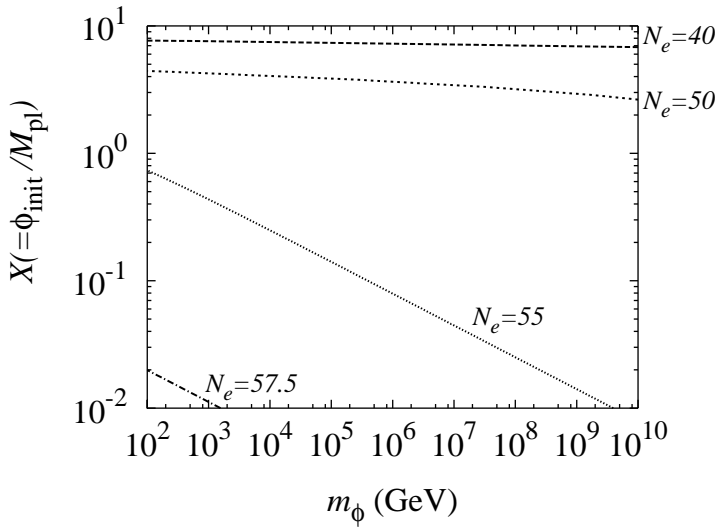}
\caption{Contours of constant $N_e$ during inflation for the case with
$\alpha=4$.  The decay rate of the curvaton is assumed to be
$\Gamma_\phi = 10^{-10}$ GeV. }
\label{fig:phi4_efold}
\end{center}
\end{figure}

In Fig.~\ref{fig:phi4}, the 95 \% C.L. allowed regions are shown for
some values of $\Gamma_\phi$ in the $m_\phi$ vs.\ $X$ plane.  As
discussed in the previous section, contribution from the curvaton
fluctuation to the adiabatic fluctuation becomes large for small
$\phi_{\rm init}$, thus regions with small $X$ become allowed. On the
other hand, regions with large initial amplitude are excluded for most
cases. In fact, in such region, the contributions from the curvaton
fluctuation becomes large. However, with such large initial amplitude,
the second inflation happens with the energy density of the curvaton.
Accordingly the number of $e$-foldings during the first inflation
becomes smaller.  In Fig.~\ref{fig:phi4_efold}, we show contours of
constant number of $e$-foldings in this model. We can see that regions
with large $X$ give small number of $e$-foldings.  Thus the amplitude
of inflaton at the horizon crossing becomes smaller than that of usual
cases. This is the reason why regions with large $X$ cannot alleviate
the inflation model even though the contribution to the curvature
perturbation from the fluctuation of the curvaton is large there.

\subsubsection{$\alpha\geq 6$ case}

\begin{figure}[t]
\begin{center}
\includegraphics{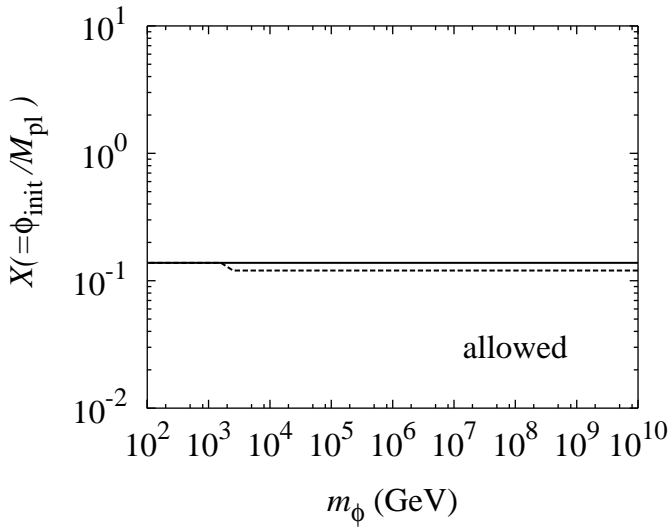}
\caption{The 95 \% C.L. allowed regions for $\alpha=6$.  The decay
rate of the inflaton is taken to be $\Gamma_\chi=10^8$ GeV.  The decay
rate of the curvaton is assumed to be $\Gamma_\phi = 10^{-18}$ GeV
(solid line) and $\Gamma_\phi = 10^{-10}$ GeV (dashed line).  The
lower side of the horizontal line is allowed for the given
$\Gamma_\chi$.}
\label{fig:phi6_1e8}
\end{center}
\end{figure}

\begin{figure}[t]
\begin{center}
\includegraphics{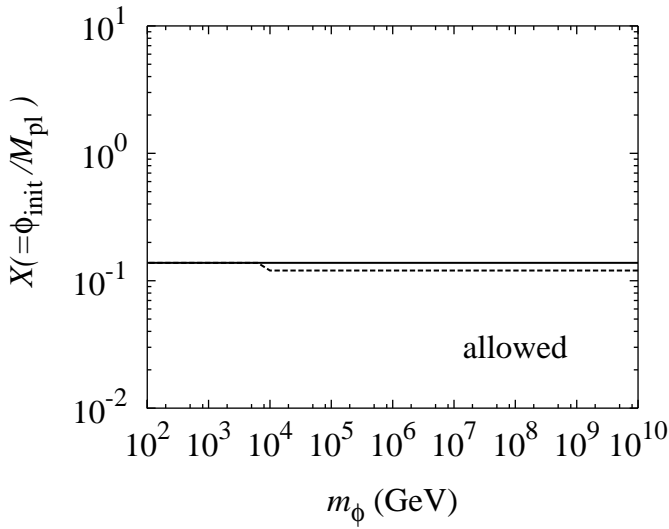}
\caption{Same as Fig.\ \ref{fig:phi6_1e8}, except for
$\Gamma_\chi=10^6\ {\rm GeV}$.}
\label{fig:phi6_1e6}
\end{center}
\end{figure}

Next we discuss the case with $\alpha=6$. The situation is very
similar to the case with $\alpha=4$.  One of differences is that, once
the inflaton starts to oscillate, its energy density evolves as
$\rho_\chi \propto a^{-4.5}$ which is stiffer than that of
radiation. Thus the constraint on the $m_\phi$ vs.\ $X$ plane depends
on the decay rate of the inflaton in this case. We present the results
for $\Gamma_\chi = 10^8$ GeV and $10^6$ GeV in
Figs.~\ref{fig:phi6_1e8} and \ref{fig:phi6_1e6},
respectively. Similarly to the case with $\alpha=4$, regions with
small $X$ can relax the constraints on the inflation model.

So far, we have seen that, for $\alpha=4$ and $6$, the chaotic
inflation models become viable when the initial amplitude of the
curvaton is small enough.  As one can expect, similar results are
obtained for larger values of $\alpha$.  We have calculated the upper
bounds on $X$ which make the chaotic inflation models with $\alpha\geq
6$ viable.  The bound is given by $X\leq 0.137$ ($0.088$) for
$\alpha=6$ ($8$).  Here, we have taken $\Gamma_\chi = 10^6$ GeV,
$m_\phi = 10^2$ GeV and $\Gamma_\phi=10^{-10}$ GeV.

Since the model with larger values of $\alpha$ predicts smaller
$n_s-1$, upper bound on $X$ becomes severer for larger value of
$\alpha$.  In particular, with the decay rates and the curvaton mass
used above, models with $\alpha\geq 10$ cannot become viable even with
the curvaton.  This is because, even if the curvaton contribution
dominates, the spectral index $n_s\simeq 1-2\epsilon\simeq 1-
\alpha/2N_e$ becomes too small to be consistent with the observations.
For $\alpha=10$, for example, $n_s-1\simeq -0.1$ with $N_e\simeq 50$,
which is excluded by the observations.  One possibility to make the
models with large $\alpha$ viable is to inclease the $e$-folding
number $N_e$ since $n_s-1$ is approximately proportional to
$N_e^{-1}$ in this case.  Larger value of $N_e$ is realized with, for
example, smaller value of $\Gamma_\chi$ or larger value of
$\Gamma_\phi$.

\subsection{Natural Inflation}

The natural inflation model \cite{Freese:1990rb,Adams:1992bn} is based
on a (psuedo-)Nambu-Goldstone (NG) field which has a potential of the
form
\begin{equation}
  V_{\rm inf} =
  \Lambda^4 \left[ 1 - \cos \left( \frac{\chi}{F} \right) \right].
\end{equation}
In this case, the slow-roll parameters are given by
\begin{equation}
  \epsilon =  
  \frac{1}{2} \left( \frac{M_{\rm pl}}{F} \right)^2 
  \frac{1}{\tan^2 (\chi/2F)},
  ~~~
  \eta = \frac{1}{2} \left( \frac{M_{\rm pl}}{F} \right)^2 
  \left[ \frac{1}{\tan^2 (\chi/2F)}  - 1 \right].
\end{equation}
The inflation ends when the slow-roll condition does not hold, namely
one of the slow-roll parameters becomes $\mathcal{O}(1)$.  The
amplitude of the inflaton at the end of inflation can be obtained by
\begin{equation}
  \tan \left( \frac{\chi_{\rm end}}{2F} \right) 
  \simeq \frac{M_{\rm pl}}{\sqrt{2} F} .
\end{equation}
The number of $e$-folding can be evaluated using the standard
technique as
\begin{equation}
  N_e = \frac{2F^2}{M_{\rm pl}^2} 
  \ln \left[ \frac{\cos (\chi_{\rm end}/2F)}
    { \cos (\chi_*/2F)} \right]
\end{equation}
After the inflation, the inflaton starts to oscillate around the
minimum of the potential. Around the minimum, the potential can be
well-approximated as a quadratic form $V_{\rm inf} \simeq (1/2)
(\Lambda^2/F)^2 \chi^2$.  Thus the energy density of oscillating
inflaton field behaves as the same as that of matter.  The natural
inflation model generally predicts a red-tilted primordial spectrum
and small tensor-scalar ratio \cite{Moroi:2000jr}.  However, in the
curvaton scenario, the spectral index can approach to the
scale-invariant value, which can alleviate the constraint on this
model.  The modification can be obtained using the formulae given in
the previous section.

\begin{figure}[t]
\begin{center}
\includegraphics{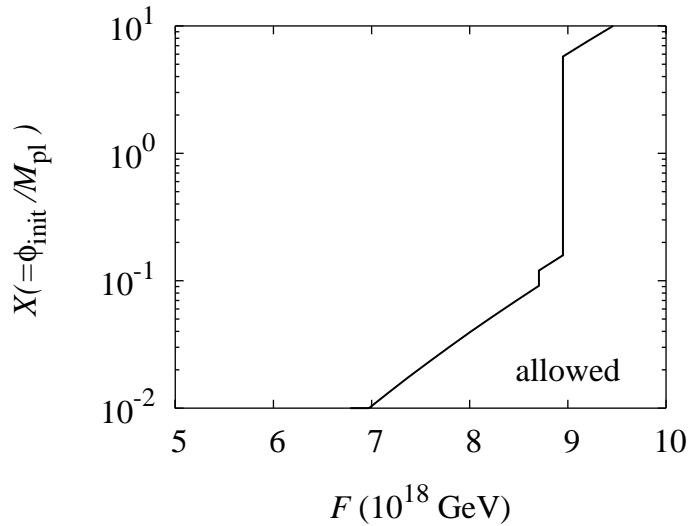}
\caption{The 95 \% C.L. allowed regions are shown for the natural
inflation model in the $F$ vs.\ $X$ plane. The lower region under the
line is allowed. The decay rate of the inflaton and the curvaton are
assumed to be $\Gamma_\chi= 10^6$ GeV and $10^{-10}$ GeV, respectively
in this figure. The mass of the curvaton is $m_\phi=10^2$ GeV.}
\label{fig:natural}
\end{center}
\end{figure}

In Fig.~\ref{fig:natural}, the 95 \% allowed regions are shown in the
$F$ vs.\ $X$ plane.  In the figure, the values of $\Lambda$ are chosen
to give a minimal value of $\chi^2$ from WMAP data using the code
provided by them \cite{Verde:2003ey}.  In principle, the constraint
also depends on the mass of the curvaton. However, since we checked
that the variation of the mass of the curvaton does not affect the
constraint much, we only show the result on the $F$ vs.\ $X$ plane.

When fluctuation of the curvaton has a small contribution to the total
curvature fluctuation, the spectral index is too small to be
consistent with the obesrvations in the natural inflation model.
However, as same as the case with the chaotic inflation models, larger
contribution from the curvaton fluctuation can liberate the model.  In
particular, the natural inflation predicts smaller spectral index as
$F$ decreases.  Correspondingly, the upper limit on $X$ where the
curvaton mechanism can liberate the model becomes smaller as it can be
seen from Fig.\ \ref{fig:natural}.

\subsection{New Inflation}

Among various possibilities of the new inflation models, we chose to
use the one proposed in \cite{Kumekawa:1994gx,Izawa:1996dv} to be
definite.  The model is based on the supersymmetric models with $Z_n$
$R$-symmetry.  We parameterize the superpotential with the parameters
$\lambda$ and $v$ as
\begin{eqnarray}
  W = \frac{\lambda}{v^{n-2}} \left(
  v^n \hat{\chi} - \frac{1}{n+1} \hat{\chi}^{n+1} \right),
  \label{eq:W_new}
\end{eqnarray}
where $\hat{\chi}$ is the superfield for the inflaton.  

Using the fact that the inflaton $\chi$ is a real scalar field, the
inflaton potential is given by\footnote
{In supergravity, extra term may arise in the inflaton potential with
non-minimal K\"ahler potential.  We assume that the higher-order terms
in the K\"ahler potential are small enough to be neglected.  It is the
case in, for example, the class of models with large cutoff scale
\cite{Ibe:2004mp}.  In addition, if we naively use the superpotential
given in Eq.\ (\ref{eq:W_new}) in supergravitity, negative
cosmological constant arises at the minimum of the potential.  We
assume that the cosmological constant problem is somehow solved since
it is not our main issue.  Thus, we assume that the inflaton potential
can be well approximated by Eq.\ (\ref{eq:V_new}) even near the
minimum of the potential.}
\begin{eqnarray}
  V_{\rm inf} = \tilde{\lambda}^2 \tilde{v}^4
  \left[ 1 - 2 \left( \frac{\chi}{\tilde{v}} \right)^{n}
    + \left( \frac{\chi}{\tilde{v}} \right)^{2n}
    \right],
  \label{eq:V_new}
\end{eqnarray}
with
\begin{eqnarray}
  \tilde{\lambda} = \frac{1}{2} \lambda,~~~
  \tilde{v} = \sqrt{2}v.
\end{eqnarray}
Here, we neglect the higher order terms which are allowed by the $Z_n$
symmetry.  Assuming that the higher-order terms are suppressed by the
inverse powers of the Planck scale, the scalar potential given above
is relevant when $\tilde{v}\ll M_{\rm pl}$.  In the following, we
first analyze the model using the above superpotential, allowing the
$v$ parameter to be as large as $M_{\rm pl}$.  After deriving the
constraints, we take account of the constraint $\tilde{v}\ll M_{\rm
pl}$ and discuss the implications.

In this model, for a given value of $\tilde{v}$, the $\tilde{\lambda}$
parameter is determined so that the cosmic density fluctuation have
the correct size.  However, as we see below, the spectral index and
the tensor-scalar ratio are independent of $\tilde{\lambda}$.

Now we evaluate the slow-roll parameters in the model.  During
inflation, the last term in Eq.\ (\ref{eq:V_new}) is irrelevant to the
dynamics of the inflaton. Thus we neglect the last term for the
moment. The slow-roll parameters are
\begin{equation}
  \epsilon =  2 n^2 \left( \frac{M_{\rm pl}}{\tilde{v}} \right)^2 
  \left( \frac{\chi_*}{\tilde{v}} \right)^{2(n-1)},
  ~~~
  \eta = - 2n(n-1) \left( \frac{M_{\rm pl}}{\tilde{v}} \right)^2 
  \left( \frac{\chi_*}{\tilde{v}} \right)^{n-2}.
\end{equation}
The number of $e$-folding is given as
\begin{equation}
  N_e = \frac{1}{2n(n-2)} \left( \frac{\tilde{v}}{M_{\rm pl}} \right)^2 
  \left( \frac{\chi_*}{\tilde{v}} \right)^{n-2}.
\end{equation}
Thus, in the standard inflationary scenario, the scalar spectral
index can be written as 
\begin{equation}
  n_s -1 = -2 \frac{n-1}{n-2} \frac{1}{N_e}.
\end{equation}
For $N_e = 50$, for example, $n_s-1$ becomes $-0.08$, $-0.06$ and
$-0.053$ for $n=3$, $4$ and $5$, respectively.  In particular, the
model with $n=3$ is marginally excluded by observations.

\begin{figure}[t]
\begin{center}
\includegraphics{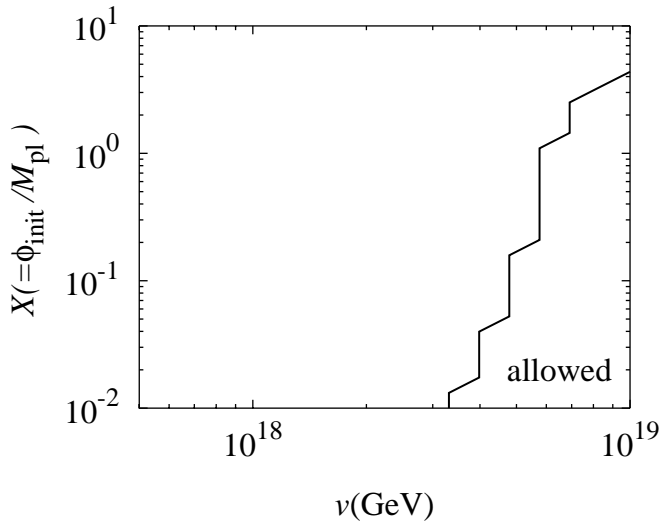}
\caption{The 95 \% C.L. allowed region is  shown for the natural
inflation model in the $v$ vs. $X$ plane. The lower region under the
line is allowed. The decay rate of the inflaton and the curvaton are
assumed to be $\Gamma_\chi= 10^6$ GeV and $10^{-10}$ GeV, respectively
in this figure. The mass of the curvaton is $m_\phi=10^2$ GeV.}
\label{fig:new}
\end{center}
\end{figure}

The situation for $n=3$ may be improved by the curvaton, as before.
In Fig.\ \ref{fig:new}, we plot the 95 \% C.L. allowed region in the
$v$ vs.\ $X$ plane. Notice that this model predicts very small value
of the $\epsilon$ parameter, in particular for the case with
$\tilde{v}\ll M_{\rm pl}$.  Using the $e$-folding number and the
amplitude of the inflaton at the horizon crossing, $\epsilon$ is given
as
\begin{equation}
  \epsilon =
  2n^2 \left( \frac{M_{\rm pl}}{\tilde{v}} \right)^2 
  \left[ \frac{1}{2n(n-2)} \left( \frac{\tilde{v}}{M_{\rm pl}} 
    \right)^2 
    \frac{1}{N_e} \right]^{\frac{2(n-1)}{n-2}}.
\end{equation} 
For $n=3$, this equation becomes $\epsilon = 18 (1/6N)^4 (
\tilde{v}/M_{\rm pl})^6$.  For $N_e = 50$ and $\tilde{v} \sim 10^{16}$
GeV, for example, the $\epsilon$ parameter becomes of order
$10^{-21}$.  Importantly, the effects of the curvaton comes in the
combination $\tilde{f}^2 \epsilon$ in the formula of the scalar
spectral index (see Eq.\ (\ref{eq:ns_mod})).  Thus, the curvaton can
help to flatten the primordial scalar power spectrum only when
$\tilde{v}$ is large enough.

As we mentioned before, the new inflation model based on the $Z_n$
$R$-symmetry requires $v\ll M_{\rm pl}$
\cite{Kumekawa:1994gx,Izawa:1996dv}.  In this case, the $\epsilon$
parameter is always very small and the curvaton contribution is always
suppressed.  Thus, it is difficult to make the (supersymmetric) new
inflation model based on $Z_n$ $R$-symmetry viable with the curvaton
mechanism.

\section{Conclusion}
\setcounter{equation}{0}

In this paper, we have studied the constraints on the inflaton
potential taking account of the effects of the curvaton.  We have
followed the basic procedure given in Ref.\ \cite{Langlois:2004nn}.
We have considered several models of the inflation for which some part
of the parameter space is excluded by the observations.  We have seen
that the constraints on these models can be relaxed once the curvaton
is introduced.

For the case of the chaotic inflation, inflation models with the
inflaton potential $V_{\rm inf}\propto\chi^{\alpha}$ can become viable
with curvaton, if $\alpha$ is small enough.  Importantly, the curvaton
contribution to the cosmic density fluctuation is inversely
proportional to $\phi_{\rm init}$ (when $\phi_{\rm init}$ is less than
$M_{\rm pl}$).  Thus, to make the models viable, upper bounds on
$\phi_{\rm init}$ is obtained.  Similar result is obtained for the
natural inflation model.

We also considered the new inflation model.  If all the parameters in
the inflaton potential (in particular, the $v$ parameter defined in
Eq.\ (\ref{eq:W_new})) are free, the curvaton mechanism can relax the
constraint on the new inflation model.  However, for some class of the
new inflation models, the $\epsilon$ parameter is supposed to be very
small.  In such a case, even the curvaton mechanism cannot liberate
the constraint.  Consequently, for the new inflation model based on
$Z_n$ symmetry proposed in \cite{Kumekawa:1994gx,Izawa:1996dv}, for
example, relatively large value of $n$ is needed.

\noindent
{\bf Acknowledgment:} 
We acknowledge the use of CMBFAST \cite{cmbfast} package for our
numerical calculations.  One of the authors (T.T.) is grateful to
Fuminobu Takahashi for useful conversation. T.T. would like to thank
the Japan Society for Promotion of Science for financial support.  The
work of T.M. is supported by the Grants-in Aid of the Ministry of
Education, Science, Sports, and Culture of Japan No.\ 15540247.

\end{document}